\documentclass[conference]{IEEEtran}
\IEEEoverridecommandlockouts
\usepackage{cite}
\usepackage{amsmath,amssymb,amsfonts}
\usepackage{algorithmic}
\usepackage{graphicx}
\usepackage{textcomp}
\usepackage{xcolor}
\usepackage{xurl}
\def\BibTeX{\rm B\kern-.05em{\sc i\kern-.025em b}\kern-.08em
    T\kern-.1667em\lower.7ex\hbox{E}\kern-.125emX}
\begin{document}

\title{Mapping Disruption Sources in the Power Grid and Implications for Resilience\\
\thanks{A part of this research was supported by the Block Center for Technology and Society of Carnegie Mellon University.}
}

\author{\IEEEauthorblockN{Maureen S. Golan}
\IEEEauthorblockA{\textit{CAEE, Cockrell School of Engineering} \\
\textit{The University of Texas at Austin}\\
Austin, USA \\
mgolan@utexas.edu}
\and
\IEEEauthorblockN{Javad Mohammadi}
\IEEEauthorblockA{\textit{CAEE, Cockrell School of Engineering} \\
\textit{The University of Texas at Austin}\\
Austin, USA \\
javadm@utexas.edu}

}

\maketitle

\begin{abstract}
Developing models and metrics that can address resilience against disruptions is vital to ensure power grid reliability and that adequate recovery and adaptation mechanisms are in place. In this paper, we propose a novel disruption mapping approach and apply it to the publicly available U.S. Department of Energy DOE-417 Electric Emergency and Disturbance Report to holistically analyze the origin of anomalous events and their propagation through the cyber, physical and human domains. We show that capturing the disruption process onset has implications for quantifying, mitigating, and reporting power grid resilience.    
\end{abstract}


\vspace{-.05in}
\section{Introduction}
\vspace{-.05in}
Utilities and regulatory authorities rely on metrics based on tracked data to assess the power grid's reliability and plan for future operating conditions. However, metrics that analyze the resilience of the power grid are immature and ill-equipped to provide utilities and regulatory authorities the information they need to ensure both reliability and resilience of the power grid in the changing context of power grid operations and development \cite{NAS2021}. Therefore, we propose a method for mapping data that utilities are required to report to different dimensions of disruption to enable a comprehensive understanding of the relationship between disruption sources and system resilience. 

\vspace{-.05in}
\subsection{Motivation}
\vspace{-.05in}
As extreme weather events and unprecedented cyber, physical, and supply chain disruptions exacerbate the changing grid landscape, resilience is a complementary concept to reliability that needs to be understood and modeled to protect the power grid. Although reliability and resilience strategies can be implemented differently by individual operators, alleviating disruption impacts requires revisiting electric grid planning, operation, maintenance and monitoring procedures\cite{Robb2021}. Leveraging resilience principles as a critical aspect of reliability can support flexible design approaches and timely recovery for differing threats and disruptions \cite{NERC2018}. Historically, reliability standards have been tracked and enforced with great precision and oversight (e.g., N-1 or N-X contingency planning under predicted scenarios, and the standardized metrics SAIDI, SAIFI, CAIDI) \cite{NERC2022,NERC2018, jin2021,MISHRA2020}. However, guidance on resilience for the power grid is in its nascency and lacks models or metrics to support effective resilience strategies \cite{DOE2019}.

\begin{figure}
\includegraphics[width=\columnwidth]{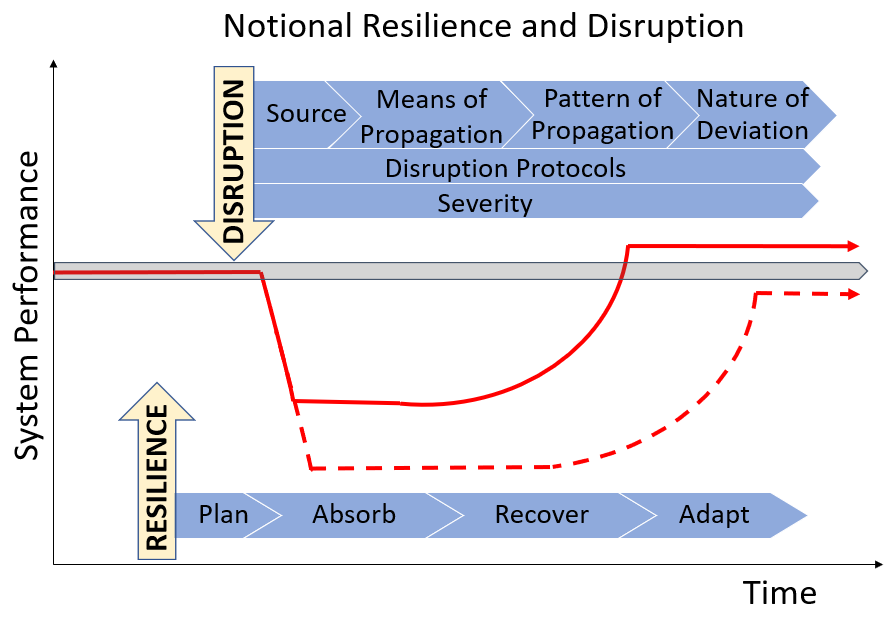}
\caption{A more resilient system (solid red line) will regain it's optimal function (grey arrow) sooner and be able to adapt better to a changing environment after a disruption than a less resilient system (dashed red line). The stages of disruption and resilience align over the resilience curve (Adapted from \cite{golan2021}).}  
\label{fig: resilience and disruption}
\end{figure}

A resilient power grid is crucial for energy security and sustainability, and reduces service restoration time and economic burdens to society and the utility \cite{Wang2017}. Resilience focuses on recovery and adaptation, expanding on risk-based metrics that use expected vulnerabilities and expected consequences to weigh the cost and benefit of hardening the system against a given hazard \cite{jin2021}. 
Resilience is commonly understood to encompass four stages – plan, absorb, recover, adapt – and emphasizes the ability of a system to recover its critical functionality as quickly as possible and with the least amount of lost function \cite{Galaitsi2021, NERC2018} (See Fig. \ref{fig: resilience and disruption}). Notably, resilience emphasizes the consequence of an outage\cite{Sandia2022}.

Another important distinction between the concepts of reliability and resilience is that reliability assumes the power grid is operating under normal (undisrupted) system conditions, whereas resilience requires a disruption to occur to be evaluated and measured \cite{Galaitsi2021}. Since resilience assumes a disrupted system state, mapping disruptions from origination across the different power grid domains is a first step towards developing a resilience model and metric(s) that can provide valuable information to system operators and developers. 

Without a systematic approach to implementing resilience across the millions of components, stakeholders, and entities spread across the globe, system managers are planning for grid failure \cite{NAS2017}. The development of comparative metrics that allow for a better understanding of the different dimensions of resilience is necessary \cite{NERC2021}. Although some studies analyze resilience, these studies often rely on hard to gather data or synthesized data sets and do not comprehensively measure resilience. In order to meet the needs of regulatory authorities, utilities, and the public, resilience metrics should be developed based on existing and already collected data and be able to track real-world disruptions from their source to impact. 

\subsection{Literature Review}
Even though many publications cover the topic of power grid resilience from different perspectives, quantitatively measuring resilience in a comprehensive and widely applicable manner remains elusive \cite{DOE2019, NERC2021}. The literature reviews of the power grid resilience field show a lack of resilience metrics that can be applied by industry and regulators, and that most studies focus on certain event types or mitigation strategies.  
For example, RAND reviewed the state of electric power resilience metrics, finding an array of metrics used to describe resilience, but no cohesive metric addressing the entire disruption timeline \cite{RAND2015}. In their review, \cite{Mar2019} analyze where mitigation approaches fall on the resilience curve and how disruption is categorized. They find that 80\%+ of power grid resilience research is based on natural causes of disruption and that blackouts and cascading failures are two of the most studied. Authors in \cite{Rickerson2019} find that the value of avoided power interruptions is used as the “standard proxy” for quantifying resilience, and that no existing methods meet criteria for regulator usefulness, defined by duration, scalability, ease of use, and output criteria. Reviews of related fields such as supply chain resilience similarly find a prevalence of proxies rather than a cohesive numeric measure of resilience \cite{golan2020trends}. 
Besides reviews, power grid resilience publications can be categorized into three topics discussed in the following subsections.

\subsubsection{Specific Components and Segments of the Power Grid}
Authors in\cite{Liu2017,Masrur2021} evaluate microgrid impact on power grid resilience from different perspectives. While \cite{Liu2017} focus on developing resilience indices (i.e., number of line outages, loss of load, expected demand not supplied, and difficulty of grid recovery), \cite{Masrur2021} study economic operations under blackout scenarios to estimate money savings over the life of the microgrid. Also, \cite{MISHRA2020} use a risk score profile based on both cyber and physical threats to examine microgrid resilience.

Others,\cite{Papic2019,Papic2020}, focus on the impact of overlapping outages, or clusters, for transmission grid elements rated at 200 kV+. Analyzing data from the North American Electric Reliability Corporation (NERC) Transmission Availability Data System (TADS), they find while lightning initiates the largest number of clusters, sustained outages are more frequently caused by failed AC substation equipment, followed by human error. Their results show that the overwhelming literature focus on natural hazards misses large sources of disruption spread.

Modeling distributed energy resources (DERs) is also an emerging resilience topic. For instance, \cite{Schadler2021} assess the ability of Germany’s grid to meet renewable penetration requirements per its 2035 carbon reduction goals. In other works, \cite{Diahovchenko2021} focus on automated switches for DER integration and improved grid resilience, and \cite{Ahrens2021} analyze control systems for smart buildings for centralized and decentralized grid services coordination. 

These papers highlight the need to focus on disruptions beyond natural causes, incorporate different power grid interactions and domains, and find ways of measuring resilience beyond risk profiles, grid capacity, and economics (i.e., proxies). Analyses on a particular grid component should result in resilience metrics that are related to the larger system and interconnected networks. 

\subsubsection{Specific Disruption Types}
Authors in \cite{Huang2017} analyze predetermined emergency and preventive resilience strategies, finding that combining both emergency and preventive topology switching has the most significant impact on enhancing power grid resilience.  
In addition, \cite{McGrath2018} models attacks on substations by simulating a substation with different security upgrades and assessing which are most effective at mitigating damage, finding that increased physical barriers and armored transformer protection are the most effective.

These studies use synthetic data sets, and leverage resilience proxies such as total cost, operating costs, and other cost functions, %\cite{Huang2017}, %\cite{Shao2017} 
load shed \cite{Huang2017}, and physical substation damage \cite{McGrath2018} without consideration to impact on output. These studies also confine their resilience analyses to specific responses and mitigation strategies, meaning their resilience outcomes are system-specific and based on preconceived disruption profiles. 

\subsubsection{System as a Whole}
Authors in \cite{Che-Castaldo2021} analyze critical risk indicators across different human and natural networks that comprise the power grid as a critical societal resource, finding that electric grid resilience cannot be defined in a closed system. Reference \cite{Mukherjee2018} uses a multi-hazard approach to develop a two-stage hybrid risk estimation model to characterize key predictors of severe weather-induced power outages. They use data-mining techniques on publicly available data on major power outages (including form DOE-417), socio-economic data, state-level climatological observations, electricity consumption trends, and land-use. Moreover, \cite{Shen2018} also use form DOE-417 and look at a combination of time between disruptions,  performance loss of each disruption, and time needed for recovery. Using data from Jan. 2002 to Aug. 2016, they evaluate trends by NERC regions to find that resilience in the NPCC region improved, likely due to increased financial support of NPCC's Compliance and Enforcement program.

Despite broadening power grid resilience to include systems thinking, interconnections and socioeconomic factors, and siting the importance of not focusing on singular scenarios or time frames, resilience metrics and quantification are not analyzed directly and are not tied to disruption propagation in the overall system \cite{Che-Castaldo2021,Mukherjee2018}. In the last study, although the authors analyze trends in disruption timelines, the findings  oversimplify disruption occurrences to power grid resilience, and make it difficult to apply resilience strategies \cite{Shen2018}. 

\subsection{Contributions}
This research analyzes the beginning stages of power grid disruptions to better understand disruption propagation and impacts on resilience. We take a systems thinking approach to developing power grid disruption categories that can then be applied to resilience analytics, and ultimately quantified for use by regulators, utilities, and other stakeholders. We expand on the existing literature by addressing all disruption types, the full disruption process, and the entire cyber-physical system, and use recent and real data. Our approach facilitates the application of effective and meaningful resilience metrics and strategies for the power grid since it corresponds to the fundamental disruption process, negating the need for resilience proxies and event-specific analytics.

\section{Mapping Disruption to Power Grid Resilience}

\subsection{Data Source}
\begin{figure}
\includegraphics[width=\columnwidth]{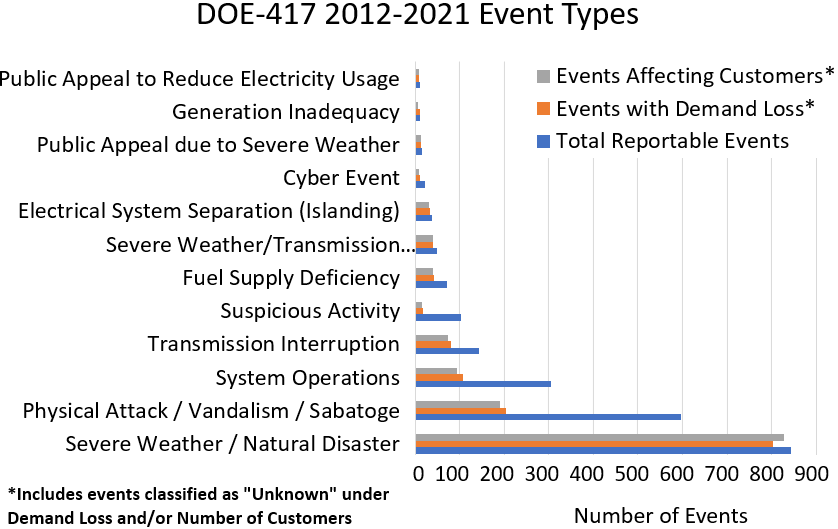}
\caption{DOE-417 Event Types cleaned and clustered into event categories for disruption mapping. Reportable events with greater than 10 reported disruptions from 2012-2021 are shown here.}  
\label{fig: Event Types}
\end{figure}

The Department of Energy’s Electric Disturbance Events form, DOE-417, is a mandatory form for utilities to file emergency incidents for national security and emergency management responsibilities \cite{DOE2022}. The form is approved by the NERC and publicly available data summaries include: date/time event began, date/time of restoration, area affected, NERC region, alert criteria, event type, demand loss, and number of customers affected. The form is generally updated every 3 years, with the last update occurring in 2021 \cite{OE4172022}. 
We use the annual summaries from 2012-2021 to analyze 2,286 incidents, grouping the “Event Types” into 46 event categories for mapping purposes. “Alert criteria” offer more disruption source details, but were only added in 2015. 
The reported events do not describe power grid resilience independently but must be associated with consequences. Given the available data, using demand loss and customers affected can begin to provide some insight into how the disruption impacts the system (See Fig. \ref{fig: Event Types}). Therefore, number of events, demand loss, and customers affected are used in our analysis and mapped to the beginning stages of disruption. Understanding the root cause of disruption is done by assuming probabilities for each propagation possibility for each event type across the disruption stages, as defined in the next section.

\subsection{Defining Disruptions}

We take a systems approach to map disruptions from the source to the ultimate consequence (see Fig. \ref{fig: resilience and disruption}). The power grid is a complex interconnected network. Thus, component-level disruptions (nodes or links) can result in network-level failures 
\cite{Durach2017,Sawik2020a,Bugert2018,Wagner2006,Tang2011,Linkov2020,golan2020trends}. In this subsection, we define major factors of a disruption process. 

\begin{itemize}
    \item \textbf{Disruption Source:} Here, we consider the origination of an adverse deviation from normal operations, i.e., the unforeseen triggering event that results in anomalies across the network. The disruption sources are broadly defined as \emph{human} and \emph{nature}. Under these sources, the scope of the disruption is classified as \emph{local} (nodes/links subject to isolated disturbances), \emph{regional} (nodes/links dispersed over an area subject to regional disturbances), and \emph{global} (nodes/links dispersed globally subject to global disturbances). We also consider the direction of disruption: \emph{supply-side/upstream} (nodes/links associated with generation through distribution; event impacting a system from within the utility), \emph{demand-side/downstream} (nodes/links associated with electricity use), and \emph{catastrophic} (force majeure; extraordinary event impacting a system from all directions).
    \item \textbf{Means of Disruption Propagation:} We account for the domain through which the violation from normal operations spreads from origination. The three domains are \emph{physical} (physical infrastructure and natural environment, including structures, processes and designs), \emph{cyber} (information and information development from physical and human domains), and \emph{human} (organizational structure, communication, and societal context of decision-making). 
    \item \textbf{Pattern of Propagation:} Characteristics of how the negative deviation from normal operations disseminates throughout the network.
    \item \textbf{Nature of Deviation:} How normal operations have been negatively impacted.
\end{itemize}

Using the above categorization, we evaluated the 2,286 reported Event Types for likelihood of stemming from human or nature, and within those, from an upstream, downstream, or catastrophic occurrence and from a local, regional, or global disruption source. We also evaluated each reported event for likelihood of propagating in three domains.

\section{Results and Discussion}

\subsection{Disruption Source}

In the source stage of a disruption, the total number of reportable events are slightly more often caused by humans. Fifty three percent of events in the scope stage and 54\% of events in the upstream/downstream stage are caused by humans. However, out of the reported events with demand and customer impacts, more of them are mapped to natural sources.\footnote{All years include “Unknown” classifications in the “Demand Loss” and “Number of Customers Affected” categories; 2014 and earlier also include “N/A” in these categories; 2013 also includes “None.” These last two entry types are note considered, whereas “Unknown” is because it does not rule out a disruption of greater than zero minutes and/or to at least one customer.} Sixty eight percent of events in the scope stage and 66\% of events in the upstream/downstream stage do not report zero demand loss. Moreover, 70\% of events in the scope stage and 68\% of events in the upstream/downstream stage do not report zero customers affected. This could be in part because human initiated events may be more visible to reporting utilities than naturally occurring events. False alarms or events that do not cause significant impacts are more likely to be reported, while naturally occurring events may only be known if an actual impact occurs to the grid. For example, Physical Attacks are also associated with suspected attacks and they may be skewed towards overreporting; whereas natural events are not necessarily known unless they cause a system interruption. Human-caused events may also have better strategies in place for absorbing impacts, eliminating system-wide disturbances, while nature-caused events may bypass the absorb phase of resilience with strategies focused on the recovery phase. Although a majority of the reported events mapped to nature result in impacts to the grid, those that are mapped to human sources still contribute to about one third of felt power grid disruptions and can offer insight into resilience strategies at different stages of the resilience curve (see Fig. \ref{fig: resilience and disruption}).  

\begin{figure}
\includegraphics[width=\columnwidth]{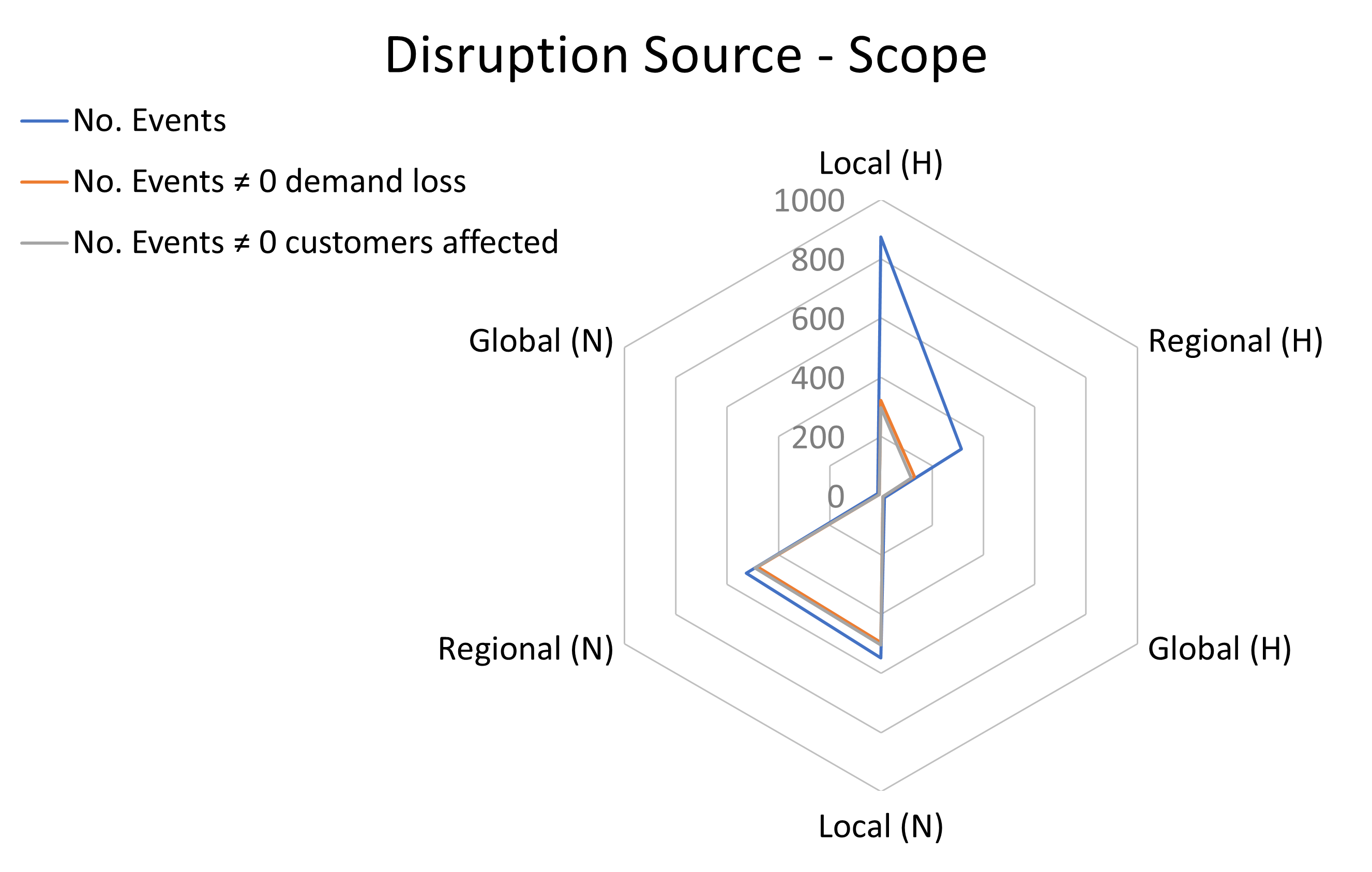}
\caption{DOE-417 event types, 2012-2021, mapped to human (H) and natural (N) sources of local, regional, and global disruption.}  
\label{fig: disruption source - scope}
\end{figure}

\begin{figure}
\includegraphics[width=\columnwidth]{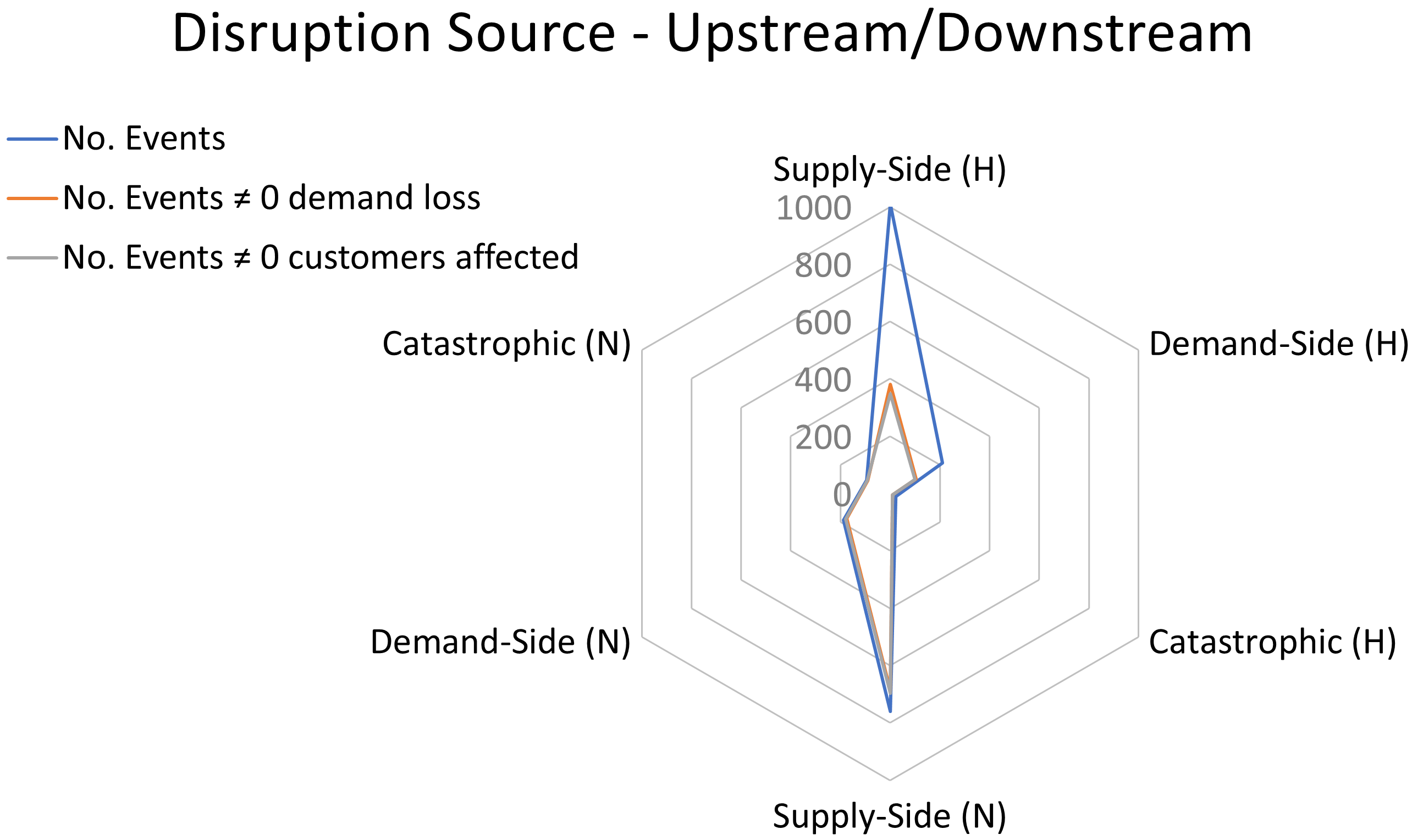}
\caption{DOE-417 event types, 2012-2021, mapped to human (H) and natural (N) sources of demand-side, supply-side, and catastrophic disruption.}
\label{fig: disruption source - supply chain}
\vspace{-.2in}
\end{figure}

Specifically under scope (see Fig. \ref{fig: disruption source - scope}), when human-caused, events tend to be more localized, whereas with natural events, they are more evenly split between local and regional sources. Under upstream/downstream (see Fig. \ref{fig: disruption source - supply chain}), both human and natural sources are more likely to be a result of upstream disruption sources. This is in part due to the fact that the reporting is done by the utility and most sources of disruption will therefore be on the power delivery side rather than the demand side. However, it is important to note that about one fifth of disruptions reporting demand or customer loss stem from downstream sources. While 17\% of the total reported events are mapped to demand-side, 19\% of total events that do not report zero demand loss and 20\% of total events that do not report zero customers impacted are mapped to demand-side impacts. This could indicate that under certain disruptive conditions, system operators may favor controlled load shedding to minimize the depth of the resilience curve – an absorb strategy – as opposed to the possibility of an uncontrolled network disruption that could impact more customers or critical customers (e.g., hospitals and fire stations).

\subsection{Means of Disruption Propagation}

\begin{figure}
\includegraphics[width=\columnwidth]{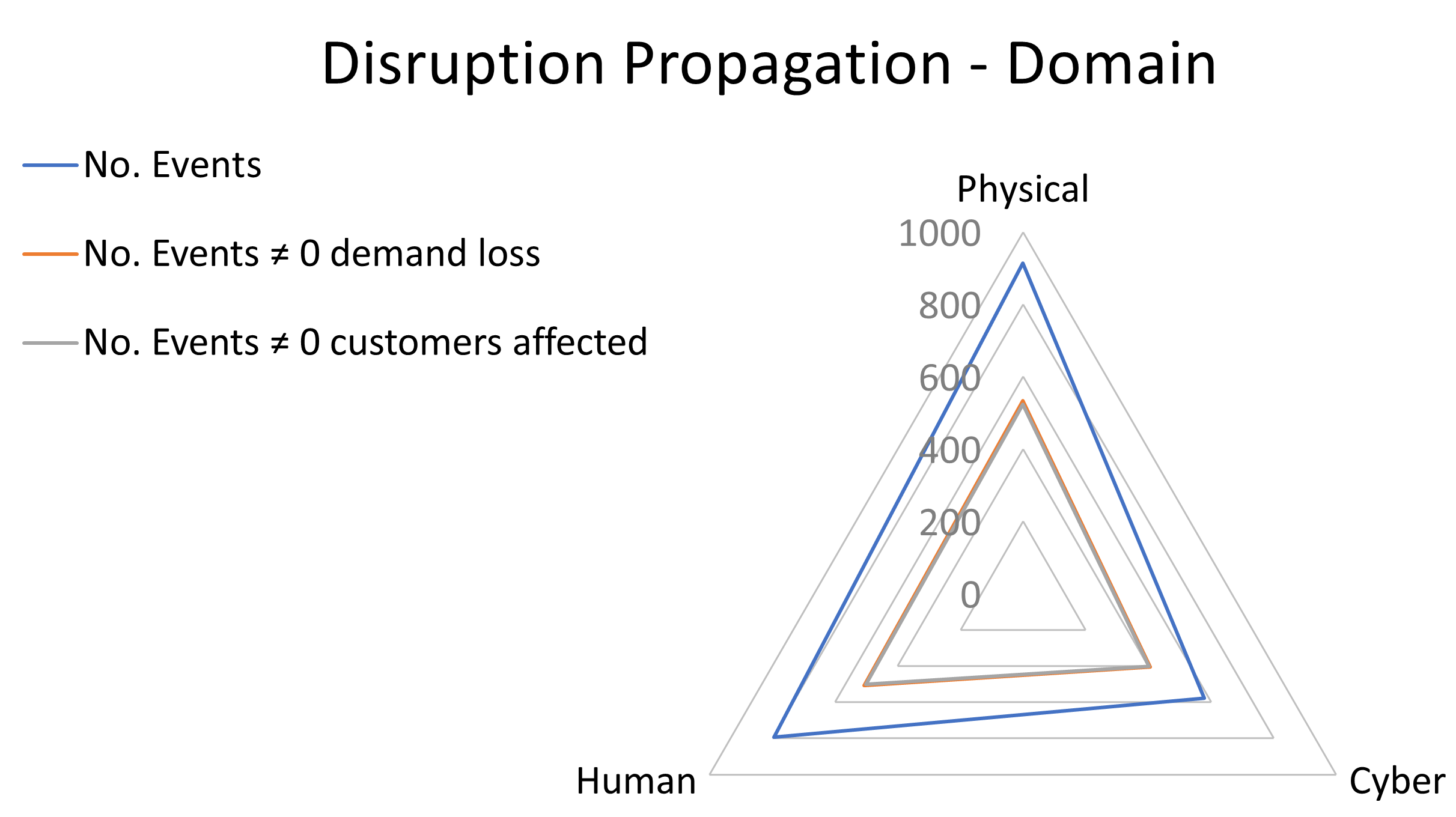}
\caption{DOE-417 event types, 2012-2021, mapped to  the domains.}
\label{fig: disruption propagation - domain}
\end{figure}

The results from mapping the disruption propagation show that once a disruption is initiated it has a fairly equal probability of impacting the power grid via all three domains. Fig. \ref{fig: disruption propagation - domain} shows that reported events propagate through cyber, physical, and human domains. This highlights the fact that the power grid is an interconnected cyber physical system and that disruptions can propagate in any domain, and likewise resilience measures should equally target all three domains. Put differently, all three must be addressed when responding to a disruptive event to minimize the resilience curve.

\subsection{Major Event Types}

\begin{figure}
\includegraphics[width=\columnwidth]{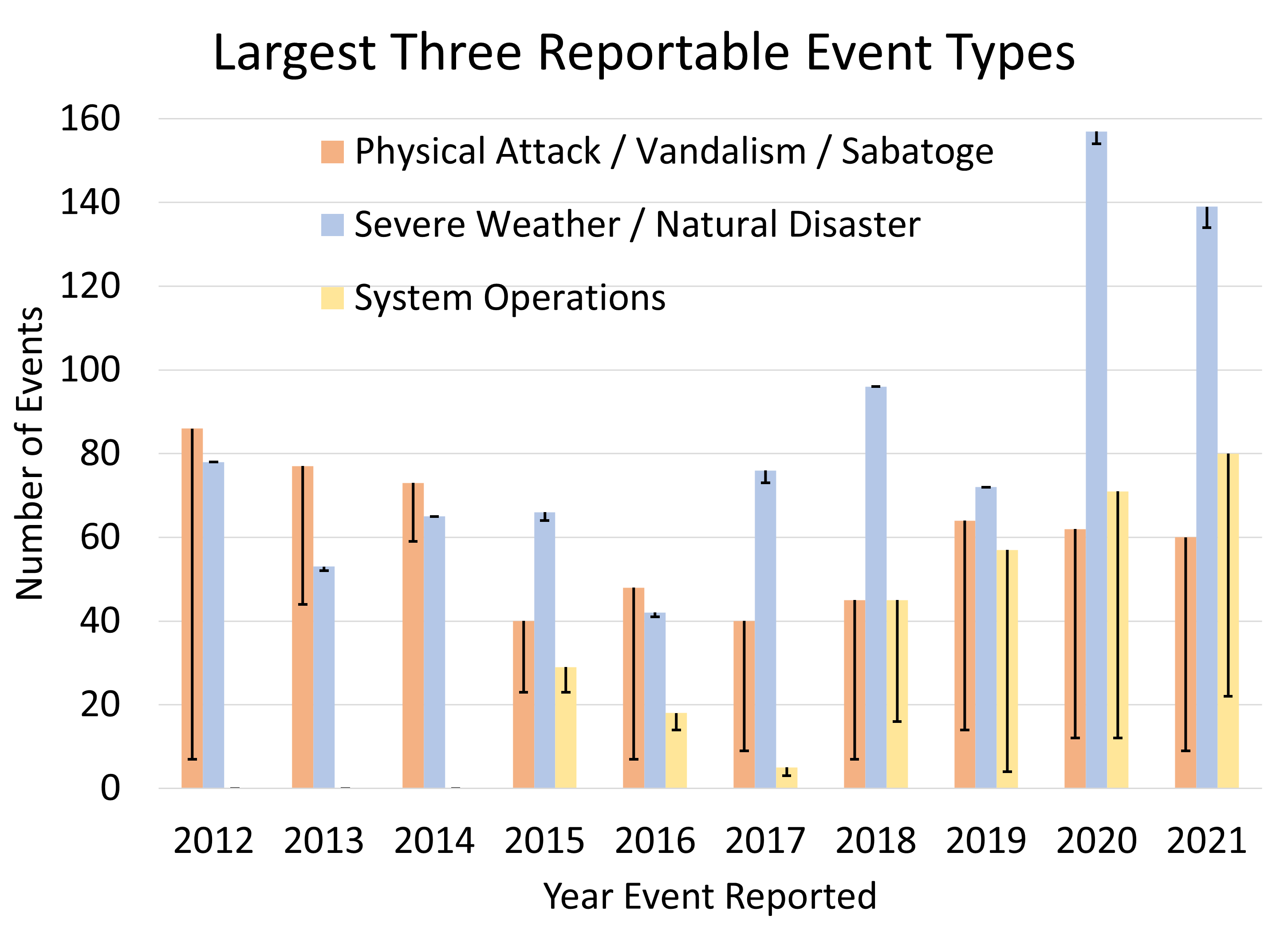}
\caption{DOE-417 event types, 2012-2021, for the largest three reportable events. The number of events with customers affected is shown by the horizontal bar overlaid on the bar graph, with the vertical line showing events that did not report affected customers. For example, there were 7 instances in 2012 where Physical Attacks resulted in a nonzero amount of customers affected. (Note that System Operations is not a reportable event until 2015.)}
\label{fig: largest 3 timeline}
\end{figure}

 Out of the 46 event categories (see Sec. II.A), three account for a majority of disruptions: Severe Weather/Natural Disaster, Physical Attack/Vandalism/Sabotage, and System Operations (See Fig. \ref{fig: Event Types}). These categories account for 76\% of all reported events, 77\% of events that do not report zero demand loss, and 78\% of events that do not report zero customers affected. Physical Attacks have been a relatively steady concern since 2012, and as such (and in addition to the discussion above) may be more likely to be prepared for, absorbed, recovered, and adapted to before customers can be affected (see Fig. \ref{fig: largest 3 timeline}). Although System Operations has only been reported since 2015 as an event type, it already comprises a significant amount of reported events. The Severe Weather category has also seen an increase in recent years, likely stemming from climate change and vulnerabilities of aging infrastructure.   

\begin{figure}
\includegraphics[width=\columnwidth]{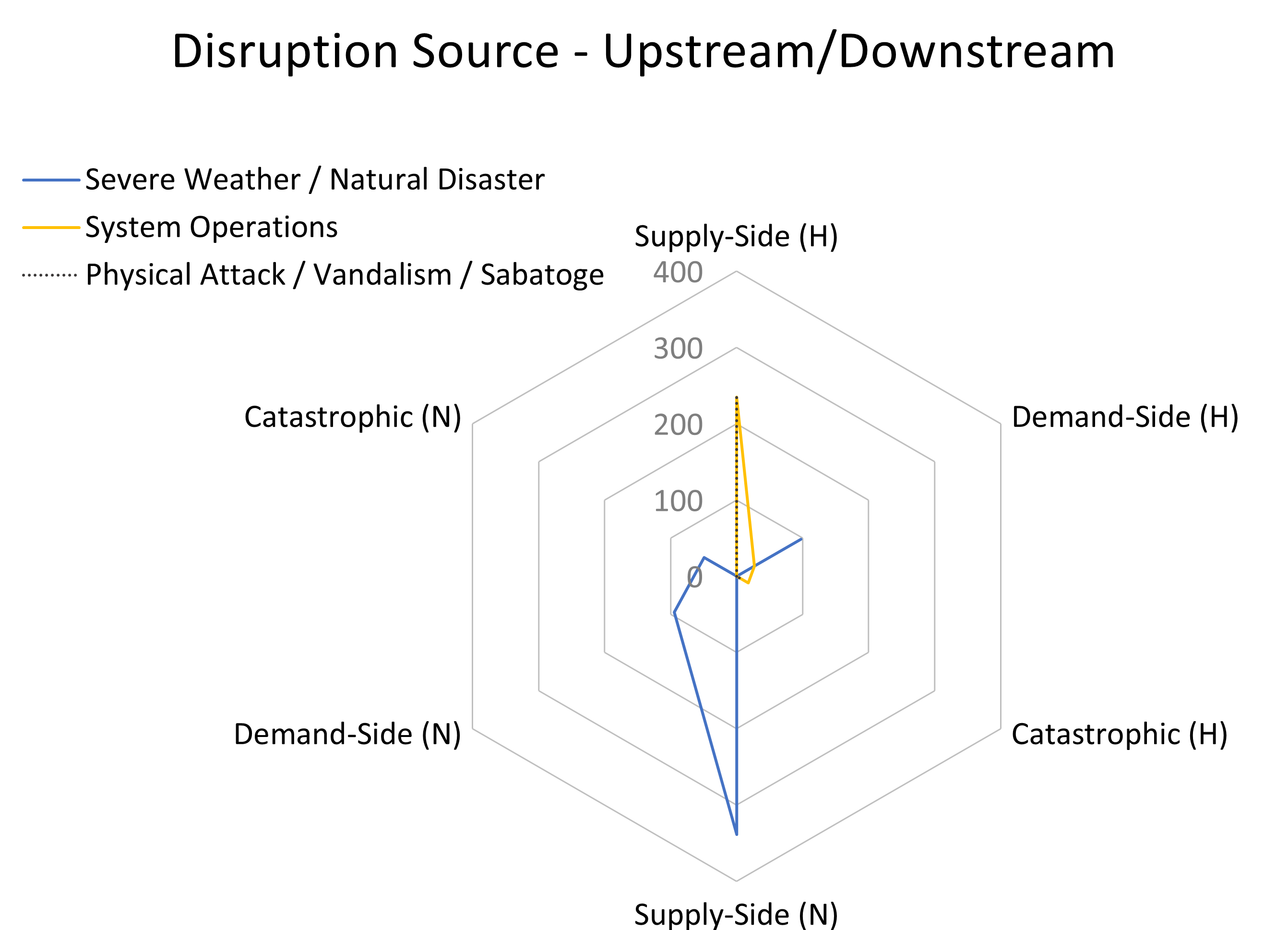}
\caption{Event categories for the largest three DOE-417 reportable events, mapped to upstream/downstream disruption sources using Alert Criteria for all reported events from 2019-2021.}
\label{fig: upstreamdownstream}
\end{figure}

These largest three categories were also mapped using the Alert Criteria for each reported event in order to gain a more detailed view at their early stages of disruption. This was done for the years 2019-2021. Individually mapping the largest three categories shows that understanding the various event types up front can signal how an event may propagate through the network. For example, analyzing the disruption source shows that Physical Attacks and System Operations are concentrated on the human-caused supply-side, while Severe Weather is concentrated on the nature-caused supply-side, with the notable exception of a concentration of human-caused demand-side disruption sources (see Fig. \ref{fig: upstreamdownstream}). This is largely attributed to the 41 instances of load shedding under emergency operational policies and the 56 instances of public appeal to reduce the use of electricity for purposes of maintaining the continuity of the Bulk Electric System.

This major event mapping was also used for a sensitivity analysis that compared   mapping from the Alert Criteria and Event Type. The largest inconsistency was within the System Operations event category, which when using Event Types overestimated the human-upstream source of disruption for events with nonzero demand loss and nonzero customers affected. This is largely due to two Alert Criteria under System Operations that provided more detail on the direction of the disruption: load shedding and public appeal. In general, the sensitivity analysis shows that although the Alert Criteria can give more insight into the disruption process for some of the broader Event Types (i.e., System Operations, Physical Attack), the results of the Event Type mapping for the larger dataset are based on valid assumptions. Ultimately, however, if generalizations can be taken out of the Event Types (as in the Severe Weather category, which had the least error and the most individual Event Types), a more detailed, accurate disruption process can be leveraged in resilience quantification.

\section{Conclusions and Future Work}
In this paper, we showcase the effectiveness of mapping the beginning stages of disruption for analyzing threats to the power grid's resilience. By segmenting disruption sources into possible starting points and propagation domains, we show that the power grid responds differently to different threats, which should be considered when assessing and strategizing resilience. We find that many reported human-caused events neither cause demand loss nor impact customers, signifying that the power grid is better at absorbing human-caused than nature-caused shocks. Adapting resilience strategies for nature-caused disruptions to similarly absorb more shocks could improve the resilience of the power grid. Moreover, downstream sources of disruption may provide a strategy for a more controlled resilience curve. We also suggest modifying required reporting to include more delineated event types and anticipate the proposed changes will help utilities and regulators better analyze disruptions and resilience. 

The fact that our mapping stems from probabilities based on assumptions derived from the DOE definitions limits our study. It is also not guaranteed that utility operators consistently report events or that the definitions adequately capture the nature of the events. We also based our analysis on the number of events (due to data limitations), which does not capture the magnitude of disruption. To better understand and quantify resilience, our future works will incorporate the magnitude and the later stages of disruption to enable mapping the entire disruption process. 
Further developing models and metrics that can track real-world disruptions is vital to adequately capture the resilience curve and ensure that the power grid meets society's needs despite unprecedented disruptions. Finding ways to use existing reporting and oversight mechanisms to gain a systems view of the power grid is essential.

\bibliographystyle{IEEEtran}
\bibliography{ref.bib}

\begin{thebibliography}{10}
\providecommand{\url}[1]{#1}
\csname url@samestyle\endcsname
\providecommand{\newblock}{\relax}
\providecommand{\bibinfo}[2]{#2}
\providecommand{\BIBentrySTDinterwordspacing}{\spaceskip=0pt\relax}
\providecommand{\BIBentryALTinterwordstretchfactor}{4}
\providecommand{\BIBentryALTinterwordspacing}{\spaceskip=\fontdimen2\font plus
\BIBentryALTinterwordstretchfactor\fontdimen3\font minus
  \fontdimen4\font\relax}
\providecommand{\BIBforeignlanguage}[2]{{%
\expandafter\ifx\csname l@#1\endcsname\relax
\typeout{** WARNING: IEEEtran.bst: No hyphenation pattern has been}%
\typeout{** loaded for the language `#1'. Using the pattern for}%
\typeout{** the default language instead.}%
\else
\language=\csname l@#1\endcsname
\fi
#2}}
\providecommand{\BIBdecl}{\relax}
\BIBdecl

\bibitem{NAS2021}
\BIBentryALTinterwordspacing
\relax{National Academies of Sciences Engineering and Medicine}, \emph{The
  Future of Electric Power in the United States}.\hskip 1em plus 0.5em minus
  0.4em\relax Washington, DC: The {N}ational {A}cademies {P}ress, 2021.
  [Online]. Available:
  \url{https://nap.nationalacademies.org/catalog/25968/the-future-of-electric-power-in-the-united-states}
\BIBentrySTDinterwordspacing

\bibitem{Robb2021}
\BIBentryALTinterwordspacing
J.~B. Robb. Reliability, resiliency, and affordability of electric service in
  the united states amid the changing energy mix and extreme weather events.
  \uppercase{T}estimony before the \uppercase{C}ommittee on \uppercase{E}nergy
  and \uppercase{N}atural \uppercase{R}esources, \uppercase{U.S. S}enate,
  \uppercase{W}ashington \uppercase{D.C. M}arch 11, 2021. [Online]. Available:
  \url{https://www.nerc.com/news/Headlines\%20DL/NERC\%20Reliability\%20Hearing\%20Testimony\%203-11-21\%20-\%20Final.pdf}
\BIBentrySTDinterwordspacing

\bibitem{NERC2018}
\BIBentryALTinterwordspacing
NERC. {R}eliability {I}ssues {S}teering {C}ommittee: \uppercase{R}eport on
  {R}esilience. \uppercase{A}tlanta, \uppercase{G}eorgia, 2018. [Online].
  Available:
  \url{https://www.nerc.com/comm/RISC/Related\%20Files\%20DL/RISC\%20Resilience\%20Report_Approved_RISC_Committee_November_8_2018_Board_Accepted.pdf}
\BIBentrySTDinterwordspacing

\bibitem{NERC2022}
\BIBentryALTinterwordspacing
------. Reliability standards for the bulk electric systems of north america.
  \uppercase{A}tlanta, \uppercase{G}eorgia, 2022. [Online]. Available:
  \url{https://www.nerc.com/pa/Stand/Reliability\%20Standards\%20Complete\%20Set/RSCompleteSet.pdf}
\BIBentrySTDinterwordspacing

\bibitem{jin2021}
A.~S. Jin and et~al., ``Building resilience will require compromise on
  efficiency,'' \emph{Nature Energy}, vol.~6, no.~11, pp. 997--999, 2021.

\bibitem{MISHRA2020}
\BIBentryALTinterwordspacing
S.~Mishra, K.~Anderson, B.~Miller, K.~Boyer, and A.~Warren, ``Microgrid
  resilience: A holistic approach for assessing threats, identifying
  vulnerabilities, and designing corresponding mitigation strategies,''
  \emph{Applied Energy}, vol. 264, p. 114726, 2020. [Online]. Available:
  \url{https://www.sciencedirect.com/science/article/pii/S0306261920302385}
\BIBentrySTDinterwordspacing

\bibitem{DOE2019}
\BIBentryALTinterwordspacing
\relax{Department of Energy}. North american energy resilience model.
  \uppercase{O}ffice of \uppercase{E}lectricity. \uppercase{W}ashington,
  \uppercase{D.C.} [Online]. Available:
  \url{https://www.energy.gov/sites/prod/files/2019/07/f65/NAERM_Report_public_version_072219_508.pdf}
\BIBentrySTDinterwordspacing

\bibitem{Wang2017}
J.~Wang and H.~Gharavi, ``Power grid resilience [scanning the issue],''
  \emph{Proc. of IEEE}, vol. 105, no.~7, pp. 1199--1201, 2017.

\bibitem{Galaitsi2021}
\BIBentryALTinterwordspacing
S.~E. Galaitsi, J.~M. Keisler, B.~D. Trump, and I.~Linkov, ``The need to
  reconcile concepts that characterize systems facing threats,'' \emph{Risk
  {A}nalysis}, vol.~41, no.~1, pp. 3--15, 2021. [Online]. Available:
  \url{https://onlinelibrary.wiley.com/doi/abs/10.1111/risa.13577}
\BIBentrySTDinterwordspacing

\bibitem{Sandia2022}
\BIBentryALTinterwordspacing
\relax{Sandia National Laboratories}. (2022) Grid resilience. [Online].
  Available:
  \url{https://energy.sandia.gov/programs/electric-grid/resilient-electric-infrastructures/}
\BIBentrySTDinterwordspacing

\bibitem{NAS2017}
\BIBentryALTinterwordspacing
\relax{National Academies of Sciences Engineering and Medicine},
  \emph{Enhancing the Resilience of the Nation's Electricity System}.\hskip 1em
  plus 0.5em minus 0.4em\relax Washington, DC: The National Academies Press,
  2017. [Online]. Available:
  \url{https://nap.nationalacademies.org/catalog/24836/enhancing-the-resilience-of-the-nations-electricity-system}
\BIBentrySTDinterwordspacing

\bibitem{NERC2021}
\BIBentryALTinterwordspacing
NERC. 2021 state of reliability: An assessment of 2020 bulk power system
  performance. [Online]. Available:
  \url{https://www.nerc.com/pa/RAPA/PA/Performance\%20Analysis\%20DL/NERC_SOR_2021.pdf}
\BIBentrySTDinterwordspacing

\bibitem{golan2021}
M.~S. Golan, B.~D. Trump, J.~C. Cegan, and I.~Linkov,
  ``\BIBforeignlanguage{eng}{The vaccine supply chain: A call for resilience
  analytics to support covid-19 vaccine production and distribution},'' in
  \emph{\BIBforeignlanguage{eng}{COVID-19: Systemic Risk and Resilience}}, ser.
  Risk, Systems and Decisions.\hskip 1em plus 0.5em minus 0.4em\relax Cham:
  Springer International Publishing, 2021, pp. 389--437.

\bibitem{RAND2015}
\BIBentryALTinterwordspacing
H.~H. Willis and K.~Loa, \emph{Measuring the Resilience of Energy Distribution
  Systems}.\hskip 1em plus 0.5em minus 0.4em\relax Santa Monica, CA: RAND
  Corporation, 2015. [Online]. Available:
  \url{https://www.rand.org/pubs/research_reports/RR883.html}
\BIBentrySTDinterwordspacing

\bibitem{Mar2019}
A.~Mar, P.~Pereira, and J.~F.~Martins, ``\BIBforeignlanguage{eng}{A survey on
  power grid faults and their origins: A contribution to improving power grid
  resilience},'' \emph{\BIBforeignlanguage{eng}{Energies}}, vol.~12, no.~24,
  pp. 4667--, 2019.

\bibitem{Rickerson2019}
\BIBentryALTinterwordspacing
W.~Rickerson, J.~Gillis, and M.~Bulkeley, ``The value of resilience for
  distributed energy resources: An overview of current analytical practices.
  \uppercase{R}eport prepared for the \uppercase{N}ational
  \uppercase{A}ssociation of \uppercase{R}egulatory \uppercase{U}tility
  \uppercase{C}ommissioners \uppercase{(NARUC)},'' Tech. Rep., 04 2019.
  [Online]. Available:
  \url{https://pubs.naruc.org/pub/531AD059-9CC0-BAF6-127B-99BCB5F02198}
\BIBentrySTDinterwordspacing

\bibitem{golan2020trends}
M.~S. Golan, L.~H. Jernegan, and I.~Linkov, ``Trends and applications of
  resilience analytics in supply chain modeling: systematic literature review
  in the context of the covid-19 pandemic,'' \emph{Environment Systems and
  Decisions}, vol.~40, no.~2, pp. 222--243, 2020.

\bibitem{Liu2017}
X.~Liu, M.~Shahidehpour, Z.~Li, X.~Liu, Y.~Cao, and Z.~Bie,
  ``\BIBforeignlanguage{eng}{Microgrids for enhancing the power grid resilience
  in extreme conditions},'' \emph{\BIBforeignlanguage{eng}{IEEE {T}rans. on
  {S}mart {G}rid}}, vol.~8, no.~2, pp. 589--597, 2017.

\bibitem{Masrur2021}
H.~Masrur, A.~Sharifi, M.~R. Islam, M.~A. Hossain, and T.~Senjyu,
  ``\BIBforeignlanguage{eng}{Optimal and economic operation of microgrids to
  leverage resilience benefits during grid outages},''
  \emph{\BIBforeignlanguage{eng}{International {J}ournal of {E}lectrical
  {P}ower 'I\&' {E}nergy {S}ystems}}, vol. 132, no.~C, pp. 107\,137--, 2021.

\bibitem{Papic2019}
M.~Papic, S.~Ekisheva, J.~Robinson, and B.~Cummings, ``Multiple outage
  challenges to transmission grid resilience,'' in \emph{2019 IEEE Power 'I\&'
  Energy Society General Meeting (PESGM)}, 2019, pp. 1--5.

\bibitem{Papic2020}
M.~Papic, S.~Ekisheva, and E.~Cotilla-Sanchez, ``\BIBforeignlanguage{eng}{A
  risk-based approach to assess the operational resilience of transmission
  grids},'' \emph{\BIBforeignlanguage{eng}{Applied {S}ciences}}, vol.~10,
  no.~14, pp. 4761--, 2020.

\bibitem{Schadler2021}
Y.~Schädler, M.~Sorg, and A.~Fischer, ``\BIBforeignlanguage{eng}{Measurement
  data‐driven investigation of the actual power grid resilience with
  increasing renewable energy feed‐in},''
  \emph{\BIBforeignlanguage{eng}{Energy science 'I\&' engineering}}, vol.~10,
  no.~1, pp. 145--154, 2022.

\bibitem{Diahovchenko2021}
I.~M. Diahovchenko, G.~Kandaperumal, A.~K. Srivastava, Z.~I. Maslova, and S.~M.
  Lebedka, ``\BIBforeignlanguage{eng}{Resiliency-driven strategies for power
  distribution system development},'' \emph{\BIBforeignlanguage{eng}{Electric
  {P}ower {S}ystems {R}esearch}}, vol. 197, pp. 107\,327--, 2021.

\bibitem{Ahrens2021}
M.~Ahrens, F.~Kern, and H.~Schmeck, ``\BIBforeignlanguage{eng}{Strategies for
  an adaptive control system to improve power grid resilience with smart
  buildings},'' \emph{\BIBforeignlanguage{eng}{Energies}}, vol.~14, no.~15, pp.
  4472--, 2021.

\bibitem{Huang2017}
G.~Huang, J.~Wang, C.~Chen, J.~Qi, and C.~Guo,
  ``\BIBforeignlanguage{eng}{Integration of preventive and emergency responses
  for power grid resilience enhancement},'' \emph{\BIBforeignlanguage{eng}{IEEE
  Trans. Power Systems}}, vol.~32, no.~6, pp. 4451--4463, 2017.

\bibitem{McGrath2018}
J.~McGrath, ``\BIBforeignlanguage{eng}{Will updated electricity infrastructure
  security protect the grid? a case study modeling electrical substation
  attacks},'' \emph{\BIBforeignlanguage{eng}{Infrastructures}}, vol.~3, no.~4,
  pp. 53--, 2018.

\bibitem{Che-Castaldo2021}
J.~P. Che-Castaldo and et~al., ``\BIBforeignlanguage{eng}{Critical risk
  indicators (cris) for the electric power grid: a survey and discussion of
  interconnected effects},'' \emph{\BIBforeignlanguage{eng}{Environment
  {S}ystems 'I\&' {D}ecisions}}, vol.~41, no.~4, pp. 594--615, 2021.

\bibitem{Mukherjee2018}
S.~Mukherjee, R.~Nateghi, and M.~Hastak, ``\BIBforeignlanguage{eng}{A
  multi-hazard approach to assess severe weather-induced major power outage
  risks in the u.s},'' \emph{\BIBforeignlanguage{eng}{Reliability {E}ngineering
  'I\&' {S}ystem {S}afety}}, vol. 175, pp. 283--305, 2018.

\bibitem{Shen2018}
L.~Shen, B.~Cassottana, and L.~C. Tang, ``\BIBforeignlanguage{eng}{Statistical
  trend tests for resilience of power systems},''
  \emph{\BIBforeignlanguage{eng}{Reliability {E}ngineering 'I\&' {S}ystem
  {S}afety}}, vol. 177, pp. 138--147, 2018.

\bibitem{DOE2022}
\BIBentryALTinterwordspacing
\relax{Department of Energy}. Electric disturbance events (doe-417). [Online].
  Available: \url{https://www.oe.netl.doe.gov/oe417.aspx}
\BIBentrySTDinterwordspacing

\bibitem{OE4172022}
\relax{OE-417 Help Desk}, email communication, 2022.

\bibitem{Durach2017}
C.~F. Durach, P.~C. Glasen, and F.~Straube,
  ``\BIBforeignlanguage{eng}{Disruption causes and disruption management in
  supply chains with chinese suppliers: Managing cultural differences},''
  \emph{\BIBforeignlanguage{eng}{Int. {J}ournal of {P}hysical {D}istribution
  'I\&' {L}ogistics {M}anagement}}, vol.~47, no.~9, pp. 843--863, 2017.

\bibitem{Sawik2020a}
T.~Sawik, ``\BIBforeignlanguage{eng}{Selection of resilient multi-tier supply
  portfolio},'' in \emph{\BIBforeignlanguage{eng}{Supply Chain Disruption
  Management}}, ser. International Series in Operations Research 'I\&'
  Management Science.\hskip 1em plus 0.5em minus 0.4em\relax Cham: Springer
  International Publishing, 2020, pp. 367--400.

\bibitem{Bugert2018}
N.~Bugert and R.~Lasch, ``\BIBforeignlanguage{eng}{Supply chain disruption
  models: A critical review},'' \emph{\BIBforeignlanguage{eng}{Logistics
  {R}esearch}}, vol.~11, no.~1, pp. 1--35, 2018.

\bibitem{Wagner2006}
S.~M. Wagner and C.~Bode, ``\BIBforeignlanguage{eng}{An empirical investigation
  into supply chain vulnerability},'' \emph{\BIBforeignlanguage{eng}{Journal of
  {P}urchasing and {S}upply {M}anagement}}, vol.~12, no.~6, pp. 301--312, 2006.

\bibitem{Tang2011}
O.~Tang and S.~Nurmaya~Musa, ``\BIBforeignlanguage{eng}{Identifying risk issues
  and research advancements in supply chain risk management},''
  \emph{\BIBforeignlanguage{eng}{International journal of production
  economics}}, vol. 133, no.~1, pp. 25--34, 2011.

\bibitem{Linkov2020}
I.~Linkov and et~al., ``\BIBforeignlanguage{eng}{The case for value chain
  resilience},'' \emph{\BIBforeignlanguage{eng}{Management {R}esearch {N}ews}},
  vol.~43, no.~12, 2020.

\end{thebibliography}

\end{document}